\documentclass[11pt]{article}

\usepackage[paper=a4paper,margin=1in]{geometry}
\usepackage{latexsym, graphicx}
\usepackage{amsmath, amssymb, bm, longtable, array, tabularx, subcaption}
\usepackage{appendix}
\usepackage{verbatim}
\usepackage{cite}
\usepackage[colorlinks=false]{hyperref}
\usepackage{placeins}

\usepackage{xcolor}

\newcommand{\IR}{\mathbb{R}}

\newcommand\citep{\cite}
\newcommand\be{\begin{equation}}
\newcommand\ee{\end{equation}}
\newcommand\bea{\begin{eqnarray}}
\newcommand\eea{\end{eqnarray}}

\newcommand\eref[1]{(\ref{#1})}
\renewcommand\comment[1]{}

\begin{document}
\thispagestyle{empty}
\renewcommand{\thefootnote}{\fnsymbol{footnote}}

\begin{titlepage}

${}$\\

\begin{center}

\textbf{
{\LARGE Getting CICY High}
}

\vspace{0.3in}

\textbf{\large
Kieran Bull$^{a,c}$\footnote{\texttt{pykb@leeds.ac.uk}},
Yang-Hui He$^{b,c,d,e}$\footnote{\texttt{hey@maths.ox.ac.uk}},
Vishnu Jejjala$^{f,g}$\footnote{\texttt{vishnu@neo.phys.wits.ac.za}},
Challenger Mishra$^{h}$\footnote{\texttt{challenger.mishra@gmail.com}}\\
}

\vspace{0.2in}

${}^a$ \textit{School of Physics and Astronomy, University of Leeds, LS2 9JT, UK}
\vskip0.25cm

${}^b$ \textit{Department of Mathematics, City, University of London, EC1V 0HB, UK}
\vskip0.25cm

${}^c$ \textit{Rudolf Peierls Centre for Theoretical Physics, Clarendon Laboratory, \\
	Parks Rd, University of Oxford, OX1 3PU, UK}
\vskip0.25cm

${}^d$ \textit{Merton College, University of Oxford, OX1 4JD, UK}
\vskip0.25cm

${}^e$ \textit{School of Physics, NanKai University, Tianjin, 300071, P.R.\ China}
\vskip0.25cm

${}^f$ \textit{Mandelstam Institute for Theoretical Physics, NITheP, CoE-MaSS, and School of Physics,\\
University of the Witwatersrand, Johannesburg 2050, South Africa}
\vskip0.25cm

${}^g$ \textit{David Rittenhouse Laboratory, University of Pennsylvania, Philadelphia, PA 19104, USA}
\vskip0.25cm

${}^h$ \textit{Instituto de Ciencias Matem\'aticas (ICMAT), 28049 Madrid, Spain}

\end{center}

\vspace{0.1in}

\hyphenation{CICY}
\hyphenation{CICYs}

\begin{abstract}
\noindent
Supervised machine learning can be used to predict properties of string geometries with previously unknown features.
Using the complete intersection Calabi--Yau (CICY) threefold dataset as a theoretical laboratory for this investigation, we use low $h^{1,1}$ geometries for training and validate on geometries with large $h^{1,1}$.
Neural networks and Support Vector Machines successfully predict trends in the number of K\"ahler parameters of CICY threefolds.
The numerical accuracy of machine learning improves upon seeding the training set with a small number of samples at higher $h^{1,1}$.
\end{abstract}

\end{titlepage}

\newpage

\renewcommand{\thefootnote}{\arabic{footnote}}
\setcounter{footnote}{0}

\section{Introduction}\label{intro}
Ever since Kaluza and Klein extended the original insight of Einstein, we regard the fundamental forces as having an intrinsically geometric origin.
The modern realization of this paradigm is the compactification of superstring theory down to four dimensions in order to recover the particle physics probed in experiments and inferred from astrophysical observations.
In the most straightforward approach consistent with low energy supersymmetry, the six extra dimensions predicted by string theory comprise a compact \textit{Calabi--Yau threefold}.
Geometric and topological properties of the Calabi--Yau threefold determine features of the four dimensional effective action.
For example, the Euler character of the geometry fixes the number of generations of light particles.
Starting from the work of~\cite{chew} and~\cite{gkmr}, numerous constructions of this type replicate the matter spectrum and gauge symmetries that we observe in Nature~\cite{Braun:2005ux,Braun:2005nv,Bouchard:2005ag,Anderson:2007nc,Anderson:2008uw,Anderson:2011ns,Anderson:2012yf,Anderson:2013xka}.
Na\"{\i}ve extrapolation of even the simplest class of models suggests that there are $10^{23}$ (nearly a mole's worth) of superstring derived Standard Models~\cite{Constantin:2018xkj}.

The vacuum selection problem, to find a principle that explicates which solution of the fundamental theory constitutes our world and how and why this came to be, remains an outstanding puzzle.
It is also unknown what the typical string compactification looks like and how closely this solution resembles the one we actually inhabit.
There are $7890$ complete intersection Calabi--Yau (CICY) threefolds realized as the zero locus of polynomials in complex projective space.
There are an unknown number of toric Calabi--Yau threefolds obtained from triangulation~\cite{Batyrev:1994hm,Batyrev:1994pg} of the $473\,800\,776$ reflexive polytopes in $\mathbb{R}^4$ tabulated by Kreuzer and Skarke~\cite{ks}.
Other Calabi--Yau spaces are neither CICY nor toric.
The largest available database~\cite{aghjn,cyd} describes only the toric Calabi--Yau geometries with Hodge number $h^{1,1} \le 6$.
While~\cite{hjp} explores the shape of the full Kreuzer--Skarke dataset, it suffices to notice that the distribution peaks sharply, and $910\,113$ of the polytopes sit at $(h^{1,1},h^{2,1}) = (27,27)$.
The explicit Standard Model constructions to date meanwhile correspond to geometries whose Hodge numbers are $O(1)$ rather than $O(10)$.
These are atypical as manifolds with small Hodge numbers are sparse.

Recently, a promising new approach to studying the vacuum selection problem has emerged.
The development of Big Data techniques in computer science and the broad applicability of these methods to such disparate fields as art, finance, chess and go, linguistics, medicine, music, experimental particle physics, and zoology invites us to also use these tools to investigate aspects of string phenomenology and string mathematics.
In particular, the paradigm of machine learning the landscape by using neural networks to study algebraic geometry, potentially bypassing expensive computations such as Gr\"obner bases, was proposed in~\cite{he2017deep, He:2017set} (\textit{cf.}~a pedagogical introduction in~\cite{He:2018jtw}).
Already, there has been a significant amount of work in this direction, ranging from the studies of CICY geometries to the computation of line bundle cohomologies of toric hypersurfaces~\cite{he2017deep, He:2017set, He:2018jtw,  bhjm, rudelius2018learning, krefl2017machine, klaewer2018machine, wang2018learning, ruehle2017evolving, liu2017artificial, mutter2018deep, carifio2017machine,  Carifio:2017nyb, abel2014genetic,  halverson2018computational, altman2018estimating}.
These studies have relied upon a multitude of algebro-geometric databases collected over the past few decades.
A large fraction of such machine learning aided studies of the string landscape has been through the lens of neural networks with a variety of architectures~\cite{he2017deep,He:2017set, ruehle2017evolving, bhjm, krefl2017machine,wang2018learning, liu2017artificial, rudelius2018learning, mutter2018deep, klaewer2018machine}.
A host of other techniques such as linear and logistic regression, Support Vector Machines (SVMs), and random forests, to name a few, have also been used, sometimes in conjunction with neural networks~\cite{carifio2017machine, wang2018learning, altman2018estimating,bhjm,abel2014genetic,mutter2018deep,Carifio:2017nyb}.

In our previous work~\cite{bhjm}, using the CICY threefolds as a testbed, we answered the following questions.
Given the configuration matrix which defines a CICY threefold, can machine learning techniques compute the Hodge numbers of the geometry?
Can the machine deduce whether the geometry is favorable, \textit{viz}., does the number of projective space factors in the ambient space equal $h^{1,1}$?
This property is important because such geometries accommodate the construction of stable vector bundles for string model building.
Can the machine determine which geometries enjoy discrete symmetries, which are crucial for introducing Wilson lines that break the GUT symmetry to the Standard Model group?
We find that even with $50\%$ of the data for training, neural network classifiers identify the Hodge numbers at better than $80\%$ accuracy.
We select favorability with SVMs with more than $90\%$ accuracy.
Because CICYs with discrete symmetries are relatively rare ($\sim2.5$\% of all cases)~\cite{free_syms}, correctly isolating only these geometries is a comparatively less successful effort.

Heuristically, all of these investigations unfold as follows.
We segregate the dataset into two disjoint parts: a \textit{training set} $T$ and its complement $T^c$, used for \textit{validation}.
The machine is taught the associations
\be
\{ a_1, a_2, \ldots, a_n \}\quad \longrightarrow\quad \{ b_1, b_2, \ldots, b_n \}
\ee
for elements $a_i \in T$.
Based on what it has learned about the training set, the machine then tries to determine the $b_j$ corresponding to the unseen elements $a_j \in T^c$.
The selection of the elements in $T$ is performed at random at the outset.
Since the CICY threefolds have been studied for decades starting from the work of~\cite{Candelas:1987kf}, we know what the answers are and can check how frequently the algorithms arrive at the correct result.
Choosing different training sets and repeating the experiment allows us to assign error bars to the results obtained from validation.
By increasing the size of the training set incrementally, we examine the machine's learning curve.

While this provides an unexpectedly good proof of concept, the methodology is not realistic for addressing the fundamental challenge in studying Calabi--Yau compactifications:
\textit{the difficulty of a calculation increases with the Hodge numbers and the dimension}.
This, after all, is why explicit Standard Model constructions are on manifolds with Hodge numbers of $O(1)$ and why triangulating polytopes to populate the toric Calabi--Yau database~\cite{cyd} stopped at $h^{1,1}=6$.
One estimate of the total number of triangulations of the Kreuzer--Skarke dataset is $10^{10505}$~\cite{altman2018estimating}.
While there are $10^{8}$ reflexive polytopes associated to toric Calabi--Yau threefolds, the best guess in the literature is that there are $10^{18}$ reflexive polytopes whose triangulations yield toric Calabi--Yau fourfolds relevant for F-theory model building~\cite{Scholler:2018apc}.
There may be $10^{3000}$ distinctly resolvable base geometries~\cite{Taylor:2017yqr}.
The scale of these numbers renders any systematic survey of the string landscape unfeasible.
We would therefore like to develop techniques such that the training and validation sets are different in character.
We aim to train with the easy cases and use the machine to predict solutions to harder problems for which the calculations are more intricate or where the answers could be unknown.
We want as well to measure how reliable the results are when we segregate the data in this imbalanced way.
By organizing the CICY dataset into a low $h^{1,1}$ training set and a high $h^{1,1}$ validation set, we report on progress in this effort.

The structure of this letter is as follows.
In Section~\ref{sec:two}, we review the CICY threefolds.
In Section~\ref{sec:three}, we describe the machine learning architectures we employ.
In Section~\ref{sec:four}, we present the results of our investigation, which focuses on determining $h^{1,1}$ starting from the configuration matrix as the input.
In Section~\ref{sec:disc}, we provide a brief discussion and a prospectus for future work.

\section{Complete Intersection Calabi--Yau Threefolds}\label{sec:two}
For completeness, we briefly recall the relevant geometry.
We refer the reader to~\cite{Hubsch:1992nu} for a pedagogical review and references therein to the original literature.

A Calabi--Yau manifold admits a Ricci flat K\"ahler metric.
We enforce this requirement by ensuring that the first Chern class vanishes.
The simplest example of a compact Calabi--Yau threefold is the Fermat quintic in $\mathbb{P}^4$:
\be
\sum_{\alpha=1}^5 z_\alpha^5 = 0 ~, \label{eq:quintic}
\ee
where $(z_1,\ldots,z_5)\sim \lambda(z_1,\ldots,z_5)$ are coordinates on projective space and $\lambda\in \mathbb{C}^\star$.
As~\eref{eq:quintic} is a homogeneous equation, we designate this geometry $\mathbb{P}^4(5)_{-200}$.
The subscript denotes the Euler character $\chi = 2(h^{1,1}-h^{2,1})$.
This is the prototype example of a class of geometries.
Consider the \textit{configuration matrix}
\be\label{cicyconfmat}
X = \begin{array}{c} \mathbb{P}^{n_1} \cr \vdots \cr \mathbb{P}^{n_\ell} \end{array} \left( \begin{array}{ccc} q_1^1 & \cdots & q_K^1 \cr \vdots & \ddots & \vdots \cr q_1^\ell & \cdots & q_K^\ell \end{array} \right)_\chi ~.
\ee
The zero locus of a set of homogeneous polynomials defined by the given matrix over the combined set of coordinates in the product of the projective spaces $\mathbb{P}^{n_i}$ is a \textit{complete intersection Calabi--Yau} (CICY) threefold when
\bea
&& \sum_{i=1}^\ell n_i - K = 3 ~, \\
&& \sum_{a=1}^K q_a^i = n_i + 1 ~, \quad \forall\ i\in \{1,\ldots,\ell\} ~. 
\eea
The former condition imposes the requirement that the manifold is a complete intersection threefold while the latter guarantees that $c_1 = 0$.
The simplest geometries obtained in this manner are
\be
\mathbb{P}^5(3,3)_{-144} ~, \quad
\mathbb{P}^5(4,2)_{-176} ~, \quad
\mathbb{P}^6(3,2,2)_{-144} ~, \quad
\mathbb{P}^7(2,2,2,2)_{-128} ~.
\ee
The Tian--Yau manifold is another example of a CICY threefold:
\be
\begin{array}{c} \mathbb{P}^3 \cr \mathbb{P}^3 \end{array} \left( \begin{array}{ccc} 3 & 0 & 1 \cr 0 & 3 & 1 \end{array}\right)_{-18} \qquad \Longleftrightarrow \qquad \begin{array}{ccc} a^{\alpha\beta\gamma} w_\alpha w_\beta w_\gamma &=& 0 ~, \cr b^{\alpha\beta\gamma} z_\alpha z_\beta z_\gamma &=& 0 ~, \cr c^{\alpha\beta} w_\alpha z_\beta &=& 0 ~, \end{array}
\ee
where $w$ and $z$ are homogeneous coordinates on each of the two $\mathbb{P}^3$s and $a$, $b$, $c$ are generic coefficients.

For CICY threefolds, the size of the configuration matrix $X$ is constrained.
We find that
\be
K \le N_1 + N_a + 3 ~, \qquad N_1 \le 9 ~, \qquad N_a \le 6 ~.
\ee
Here, $N_1$ counts the number of $\mathbb{P}^1$ factors and $N_a$ counts the number of other projective space factors.
There are $7890$ configuration matrices ranging in size from $1\times 1$ (the quintic) to $12\times 15$ with elements $q_a^i\in [0,5]$. In this dataset, we find $70$ distinct Euler characters $\chi\in [-200,0]$ and $266$ distinct Hodge pairs $(h^{1,1},h^{2,1})$.
The topological invariant $h^{1,1}$ counts the number of two cycles and four cycles and accounts for the K\"ahler deformations, whereas $h^{2,1}$ counts the number of three cycles and accounts for the complex structure deformations.
These are, respectively, the size and shape parameters of the geometry.
Within the set of CICY threefolds,
\be
0 \le h^{1,1} \le 19 ~, \qquad
0 \le h^{2,1} \le 101 ~.
\ee
Mirror symmetry --- invariance under the interchange $h^{1,1} \leftrightarrow h^{2,1}$ --- is not a property of the dataset.
As $\chi$ is always negative, $h^{1,1} \le h^{2,1}$ for any given CICY threefold.
The Euler character is a cubic expression in the elements of the configuration matrix.
Calculating $h^{1,1}$ and $h^{2,1}$ is conceptually straightforward but requires some care~\cite{green1987all,Candelas:2008wb,Candelas:2010ve,candelas2016hodge,constantin2017hodge,mishra2017calabi}.
One of the goals of applying machine learning to this dataset is to circumvent the necessity of studious sequence chasing.
Of the CICY threefolds, $195$ possess freely acting symmetries; $37$ different finite groups appear, ranging from $\mathbb{Z}_2$ to $\mathbb{Z}_8\rtimes H_8$~\cite{free_syms}. A number of CICY threefolds also admit non-freely acting symmetries~\cite{lukas2017discrete, candelas2018highly}.

\begin{table}[!ht]
\begin{center}
\begin{tabular}{|c||c|c|c|c|c|c|c|c|c|c|}
\hline
$h^{1,1}$ & $0$ & $1$ & $2$ & $3$ & $4$ & $5$ & $6$ & $7$ & $8$ & $9$ \cr
\hline
frequency & $22$ & $5$ & $36$ & $155$ & $425$ & $856$ & $1257$ & $1463$ & $1328$ & $1036$ \cr
\hline
$N(h^{1,1})$ & $22$ & $27$ & $63$ & $218$ & $643$ & $1499$ & $2756$ & $4219$ & $5547$ & $6583$ \cr
\hline
favorable & $0$ & $5$ & $36$ & $155$ & $425$ & $837$ & $1140$ & $1112$ & $732$ & $325$ \cr
\hline\hline
$h^{1,1}$ & $10$ & $11$ & $12$ & $13$ & $14$ & $15$ & $16$ & $17$ & $18$ & $19$ \cr
\hline
frequency & $648$ & $372$ & $161$ & $72$ & $22$ & $16$ & $1$ & $0$ & $0$ & $15$ \cr
\hline
$N(h^{1,1})$ & $7231$ & $7603$ & $7764$ & $7836$ & $7858$ & $7874$ & $7875$ & $7875$ & $7875$ & $7890$ \cr
\hline
favorable & $88$ & $16$ & $3$ & $0$ & $0$ & $0$ & $0$ & $0$ & $0$ & $0$ \cr
 \hline
\end{tabular}
\end{center}
\caption{The frequency row gives the multiplicity of $h^{1,1}$ in the CICY threefold dataset.
$N(h^{1,1})$ counts the number of CICY threefolds with Hodge number less than or equal to $h^{1,1}$.
The favorable row counts the number of favorable CICYs with a given $h^{1,1}$.}
\label{tab:freq}
\end{table}
\begin{figure}[!h]
\centering
\includegraphics[width=\linewidth]{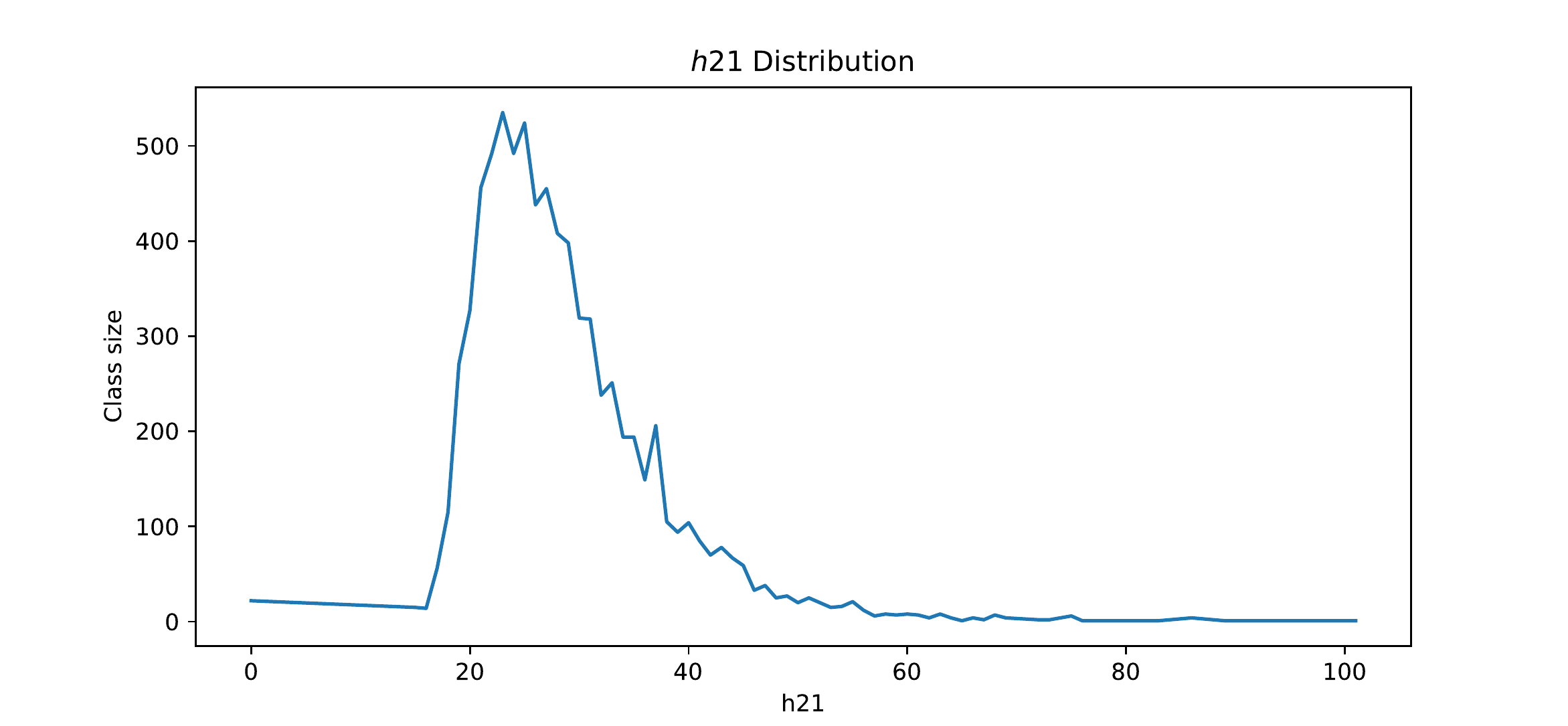}
\caption{Multiplicity of $h^{2,1}$ in CICY threefold dataset.}
\label{fig:h12dist}
\end{figure}
We tabulate the number of geometries with each value of $h^{1,1}$ in Table~\ref{tab:freq}.
Among the CICY threefolds, $4874$ out of the $7890$ are favorable, \textit{i.e.}, $h^{1,1}$ equals the number of $\mathbb{P}^n$ factors in the ambient space.
Notice that this is a slightly different definition of \textit{favorable} than others that have appeared in the literature:~\cite{Anderson:2013xka}, for example, defines a geometry as favorable when its second cohomology class descends from that of the ambient space $\mathbb{A} = \mathbb{P}^{n_1}\times\ldots\times\mathbb{P}^{n_\ell}$.
Our definition misses those geometries that can be made favorable by splitting the CICY configuration matrix further, or by
thinking of the CICY as a hypersurface in del Pezzo products.
We finally note that $h^{2,1}$ ranges over a larger interval than $h^{1,1}$.
Figure~\ref{fig:h12dist} plots the number of geometries at a given $h^{2,1}$.
Knowing $h^{1,1}$, once we compute $\chi$, the Hodge number $h^{2,1}$ is of course redundant information.
The goal of machine learning is to determine topological invariants and properties like favorability using the configuration matrix as an input.

The CICY fourfolds are catalogued in~\cite{Gray:2013mja}.
(There are $921\,497$ configuration matrices most of which correspond to elliptically fibered Calabi--Yau spaces.)
Fourfolds have four non-trivial Hodge numbers of which three are independent:
\be
4h^{1,1} - 2h^{2,1} + 4h^{3,1} - h^{2,2} + 44 = 0 ~.
\ee
Identifying all of the discrete symmetries in this dataset has not been accomplished.
Thus, there is a potential benefit to applying machine learning to this effort as well.
This is work in progress.

\section{Neural Networks and Support Vector Machines}\label{sec:three}
We briefly summarize the main ideas behind the machine learning tools we have employed in this paper (and its predecessor~\cite{bhjm}), namely, neural networks and Support Vector Machines (SVMs). Neural networks and SVMs can function as both classifiers and regressors, and as such have been the subject of active research in the machine learning community for several years.  We point the reader to the Appendices of~\cite{bhjm} for further details.
The reader familiar with these techniques can skip ahead to Section~\ref{sec:four} in which we record our results.

\subsection{Feed-forward Neural Networks}
A neural network can be thought of as a non-trivial function $f$ acting on an input vector ${v}_\text{in}$ to produce an output vector ${v}_\text{out}$, that is, $f({v}_\text{in})={v}_\text{out}$. The most successful neural network model is the \textit{feed-forward} neural network, alternately known as the \textit{multi-layer perceptron}. Architecturally, as the name indicates, a multi-layer perceptron consists of multiple layers, each of which is a collection of a number of nodes called \textit{neurons}. A multi-layer perceptron has an input layer (whose nodes correspond to the components of the input vector ${v_\text{in}}$), a number of hidden layers, and an output layer (whose nodes correspond to the components of the output vector ${v_\text{out}}$). In a feed-forward neural network, information always moves in one direction, from the input layer to the output layer. Every neuron in a given layer is connected to every neuron in the adjacent layers. Such connections are parameterized by \textit{weights} denoted by the vector ${w}$.

A single \textit{neuron} takes as input a vector ${x}$ and outputs a number $\sigma({x\cdot w}+{b})$, where $\sigma$ is the \textit{activation function}, which ordinarily maps to the range $[0,1]$. Typically $\sigma(x):=1/(1+e^{-x})$, the logistic sigmoid function, or $\text{ReLU}(x):=\text{max}(0, x)$, the rectified linear unit, or the function $\tanh$; the scalar $b$ is the \textit{bias}. The $i^{th}$ neuron in the $n^{th}$ layer, takes as input the vector $\sigma_j^{n-1}$, which are the activation values for the neurons in the previous layer, and outputs $\sigma_i^n$, where, 
\begin{equation}
\sigma_i^n := \sigma(w^n _{ij} \sigma_j^{n-1}+ b_i^n),~~\text{with~~}\sigma^0_i=({v_\text{in}})_i ~, 
\end{equation}
where $w^n_{ij}$ is the weight associated with the connection between the $i^{th}$ neuron in the $n^{th}$ layer and the $j^{th}$ neuron in the $(n-1)^{th}$ layer, and $b_i^n$ denotes the bias in the activation function for the $i^{th}$ neuron in the $n^{th}$ layer. This multilayer feedforward architecture is what allows multi-layer perceptrons to be universal approximators~\cite{hornik1989multilayer}. 

Learning happens when the multi-layer perceptron is trained to output desired vectors. Consider a multi-layer perceptron with $m$ layers being trained on a set of size $n$. For the $i^{th}$ training example, denote the output vector of the multi-layer perceptron by $\sigma^m|_i$, and the desired output vector by ${t}_i$. The mean squared error cost function can then be defined as
\begin{equation}\label{meansqerror}
E := \frac{1}{n} \sum_{i=1}^n \left({\sigma}^m|_i-{t}_i
\right)^2~.
\end{equation}
The idea is to minimize $E$ by adjusting the weights ($w^k _{ij}$) and biases ($b_i^k$) in the multi-layer perceptron. In general this is done efficiently via gradient descent using the technique of \textit{error back propagation}. 
Without going into derivations, here we simply inform the reader that the prescribed adjustments, or shifts in the values of the weights and biases for a gradient descent step, are given by
\begin{align}
	\Delta w_{ij}^k = - \eta \frac{\partial E}{\partial w_{ij}^k}\,,\quad \quad\Delta
	b_i^k = - \eta \frac{\partial E}{\partial b_i^k}\,,
\end{align}
where $\eta$ is the \textit{learning rate}. The parameter $\eta$ should be chosen judiciously since it has a strong bearing on the convergence of the gradient descent algorithm and its ability to find the true minima. Once all the weights and biases have been set by above, the neural network is trained. 

There are neural network architectures which allow for layers in which the neurons do not receive a weighted sum from all the neurons in the previous layer, but employ a \textit{kernel} (grid) that restricts the neurons that can contribute. Such neural networks are called convolutional neural networks. These are best suited to data whose inputs exhibit translation or rotational invariances, and are thus suited to problems in image recognition.%

Sometimes, the complexity of the neural network is such that it possesses more computing potential than is actually required. This leads to the problem of overfitting, wherein the accuracy of the neural network against unseen data stops improving despite growing training accuracy. The technique of dropout provides a way to counter to this, by randomly dropping neurons along with their connections from the neural network during training. This is a proven strategy against overfitting and tries to force the neurons to learn more general features of the dataset~\cite{srivastava2014dropout}. 

Optimizing the error function \eqref{meansqerror} even for a relatively simple architecture involves the tuning of a large number of weights and biases (often running into tens of thousands), which can be a drawback of neural networks. SVMs, which we discuss next, take a geometric approach to learning, and typically do not require as many parameters.

\subsection{Support Vector Machines}
Support Vector Machines (SVMs) are natural binary classifiers and are thought to be one of the best off the shelf supervised machine learning algorithms. The simplest SVM is a binary classifier for linearly separable data. The classification is performed by finding an optimal hyperplane that can separate clusters of points from the two classes in the feature space. This can be extended to tackle non-linearly separable data (using the so called \textit{kernel trick})  and data that have multiple classes~\cite{svm}. 

The simplest situation to consider is the binary classification of points in $\IR^n$. We begin by defining a hyperplane ${\cal H}$ with a normal vector $w$ by 
\begin{equation}
	{\cal H}: =\left\{ x \in \mathbb{R}^n | f({x}) := {w \cdot x} + b = 0
	\right\}.
	\label{eq:hplane_def}
\end{equation}
The idea is to find the ${\cal H}$ such that data points in the two classes lie as far from it as possible. Alternately, one maximizes the \textit{margin}, which is the distance along the normal vector $w$, between the two vectors that are the closest to the hyperplane ${\cal H}$ on either side. Such vectors are called \textit{support vectors} and it turns out that they fully specify the SVM. If we denote them by $x^\pm$, corresponding to the two classes, the margin is then given by ${\cal M}:=w\cdot (x^+-x^-)/|w|$. Since rescaling $w$ and $b$ by the same factor does not change ${\cal H}$, one can rescale $w$ such that $f(x^\pm)=\pm1$. This reduces the margin ${\cal M}$ to $2/|w|$. Thus an $x_i\in\IR^n$ is classified by the SVM using the function $\text{sign}(f(x_i))\in\{-1,1\}$, as belonging to the positive or negative class. We denote the result of classification of $x_i$, that is, $\text{sign}(f(x_i))$ as $y_i$. An alternate statement of the problem is then
\begin{align}
	&\textbf{Optimization Problem~:~~~~~~~}\text{Min    } 
	\frac{1}{2} |{w}|^2  
	\text{    subject to    }\  
	y_i({w} \cdot {x_i}+b) \geq 1.\nonumber
\end{align}
Since the objective function of the above optimization problem is convex, the solution is relatively straightforward, using standard algorithms. One can recast this problem using Lagrange multipliers as follows:
\begin{align}
	{\cal L} := \frac{1}{2} |{w}|^2 &- \sum_i 
	\Theta_i (y_i({w} \cdot {x_i} + b) - 1) ,\nonumber\\
	\frac{\partial {\cal L}}{\partial {w}} &= {w} -
	\sum_i \Theta_i y_i {x_i} \overset{!}{=}~0 , \\
	\frac{\partial {\cal L}}{\partial b} &=
	-\sum_i \Theta_i y_i \overset{!}{=}~0, \nonumber
\end{align}
which presents the
\[
    \textbf{Dual Optimization Problem~:~~~}
\begin{cases}
    \text{Min} \quad
	\frac{1}{2} \sum_{i,j} 
	\Theta_i \Theta_j y_i y_j {x_i} \cdot {x_j}
	 - \sum_j \Theta_j , &\\
\text{subject to~}\Theta_j \geq 0 \text{~~and~~} \sum_j \Theta_j y_j =0~.     & \end{cases}
\]
The classifying function is~$\text{sign}(f(x)):= \text{sign}\left( \sum_i \left( \Theta_i y_i\ {x_i \cdot x} \right) + b \right)$. It turns out that the only non-zero $\Theta_i$s correspond to the support vectors. 

In order to deal with non-linearly separable data, one could map points in the feature space to a higher dimensional feature space where the data is linearly separable. Once the optimal hyperplane ${\cal H}$ is found, one can then map back to the initial feature space. The \textit{kernel trick} implies that this is equivalent to solving the dual optimization problem after replacing the dot product ${x_i \cdot x}$ by $\mathrm{Ker}(x_i,x)$. A common form for $\mathrm{Ker}$ is the Gaussian $\text{Ker}({x_i,x}) :=	\exp \left(\frac{-|{x_i-x}|^2}{2 \sigma} \right)$, which is what we employ in this work, since it leads to the best results. 

SVMs can act as linear regressors by attempting to fit the flattest function $f(x):=w\cdot x+b$ to the data within a residue $\epsilon$. This is equivalent to the optimization problem
\begin{align}
	&\text{Min    } 
	\frac{1}{2} |{w}|^2  
	\text{~~~subject to~~} 
	-\epsilon \leq y_i - ({w \cdot x_i} + b) \leq \epsilon~,
\end{align}
since $|\nabla f|^2= |w|^2$. Similar to the case of SVM classifiers above, one can introduce Lagrange multipliers here and decise a dual version of the problem:
\begin{align}
	{\cal L} := \frac{1}{2} |{w}|^2 - &\sum_i \Theta_i (y_i-({w} \cdot {x_i} + b) + \epsilon) +\sum_i \Theta_i^\star (y_i-({w} \cdot {x_i} + b) - \epsilon)\,, \nonumber\\
	\frac{\partial {\cal L}}{\partial {w}} &= {w} -\sum_i (\Theta_i - \Theta_i^*) y_i {x_i}\overset{!}{=} 0 , \\
	\frac{\partial {\cal L}}{\partial b} &=\sum_i (\Theta_i-\Theta_i^*) y_i \overset{!}{=} 0, \nonumber
\end{align}
leads to the dual problem
\begin{align}
	\text{Min~~} \frac{1}{2} \sum_{i,j} &(\Theta_i-\Theta_i^*) (\Theta_j-\Theta_j^*) y_i y_j \ {x_i} \cdot
{x_j} +\epsilon \sum_i \left(\Theta_i+\Theta_i^* \right) + \sum_i  y_i(\Theta_i^*-\Theta_i), 
\end{align}
subject to the conditions $\Theta_i,\Theta_i^* \geq 0$ and $\sum_i (\Theta_i-\Theta_i^*) =0$. As with the SVM classifier, one can employ the \textit{kernel trick} to fit non-linear functions to the data. 

To avoid overfitting in SVMs and allowing better generalization to unseen data, one can allow a few training points to be misclassified. This has the effect of avoiding over-constraining the hyperplane ${\cal H}$, and is achieved by replacing the condition $\Theta_i\ge0$ in the dual optimization problem by $0\le \Theta \le C$, where C is called the cost variable. 
\subsection*{Architecture}
Our analysis in this paper involves a neural network regressor as well as a classifier, and an SVM regressor. The architectures for the regressors are similar to that in~\cite{bhjm}. We use the \texttt{Keras} Python package with \texttt{TensorFlow} backend to implement the neural network. The neural network consists of a $1000$ neuron later, ReLU (rectified linear unit) activation layer, $1$ neuron summation and sigmoid activation.
We use the quadratic programming Python package \texttt{Cvxopt} to solve the SVM optimization problem.
The hyperparameters are selected by hand. We employ a Gaussian kernel with $\sigma=2.74$, $C=10$, and $\epsilon=0.01$ for predicting $h^{1,1}$, and  $\sigma=3$, and no slack for the remaining experiments. 
Calculations for the regressors are performed on a Lenovo Y50 laptop, i7-4700HQ, $2.4$ GHz quad core with $16$ GB RAM. The architecture of the neural network classifier, implemented on \texttt{Mathematica} (version 11.3), consists of two Long Short-Term Memory layers with a dropout of $0.2$, each followed by a $\tanh$ and ReLU activation in sequence, and batch normalization. This is connected to two linear layers with dropout of $0.2$, each (again) followed by a $\tanh$ and ReLU activation in sequence. The final components are a linear layer, followed by a $\tanh$ and ReLU activation in sequence. Each layer has $120$ nodes. The penultimate layer of the neural network is a softmax layer. 

\section{Predicting $h^{1,1}$}\label{sec:four}
We use machine learning to compute the Hodge number $h^{1,1}$ of CICY threefolds.
Training on the configuration matrices at low $h^{1,1}$, the algorithms successfully predict trends in the distributions of Hodge numbers at higher $h^{1,1}$, but do not provide accuracy comparable to the random sampling previously studied in~\cite{bhjm}.
This is corrected by including a small selection of samples at higher $h^{1,1}$.

We set up the experiment in two parts. In the first part, we train with configuration matrices with $h^{1,1} \le x$, and test with configuration matrices with $h^{1,1} > x$. In the second part, we repeat the experiment by augmenting the training set above with $10\%$ of the configuration matrices with $h^{1,1} > x$, randomly sampled, and test using the remaining configuration matrices. We denote these two training sets by $T_x$ and $\widetilde{T}_x$ respectively. The integer bound $x$ is a tuneable parameter. In our experiments we choose $2\le x\le 10$. With reference to Table~\ref{tab:freq}, the size of the first training set $T_x$ is $N(x)$, and the size of the validation set is $7890-N(x)$. Similarly, the size of the second training set $\widetilde{T}_x$ is $\widetilde{N}(x):=N(x)+ \lfloor\frac{7890-N(x)}{10}\rfloor$, and the size of the test set is $7890-\widetilde{N}(x)$. 
Using the training set $T_x$ ($\widetilde{T_x}$), at $h^{1,1}=7$, we train with $\sim 53$\% (58\%) of the dataset while at $h^{1,1}=9$, we train with $\sim 83$\% (85\%) of the dataset.
\begin{figure}[!ht]
\centering
\includegraphics[width=0.49\linewidth]{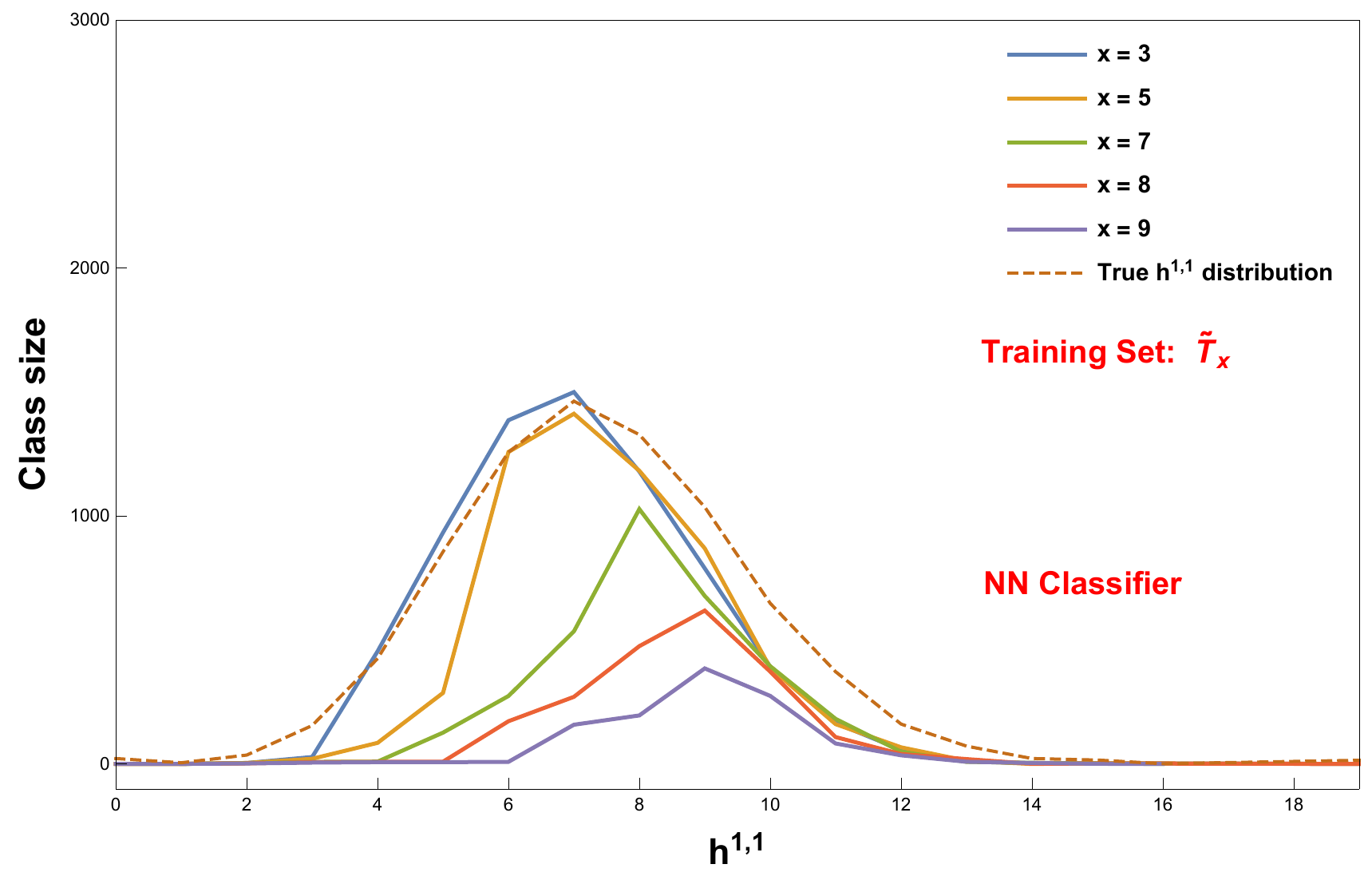}
\includegraphics[width=0.49\linewidth]{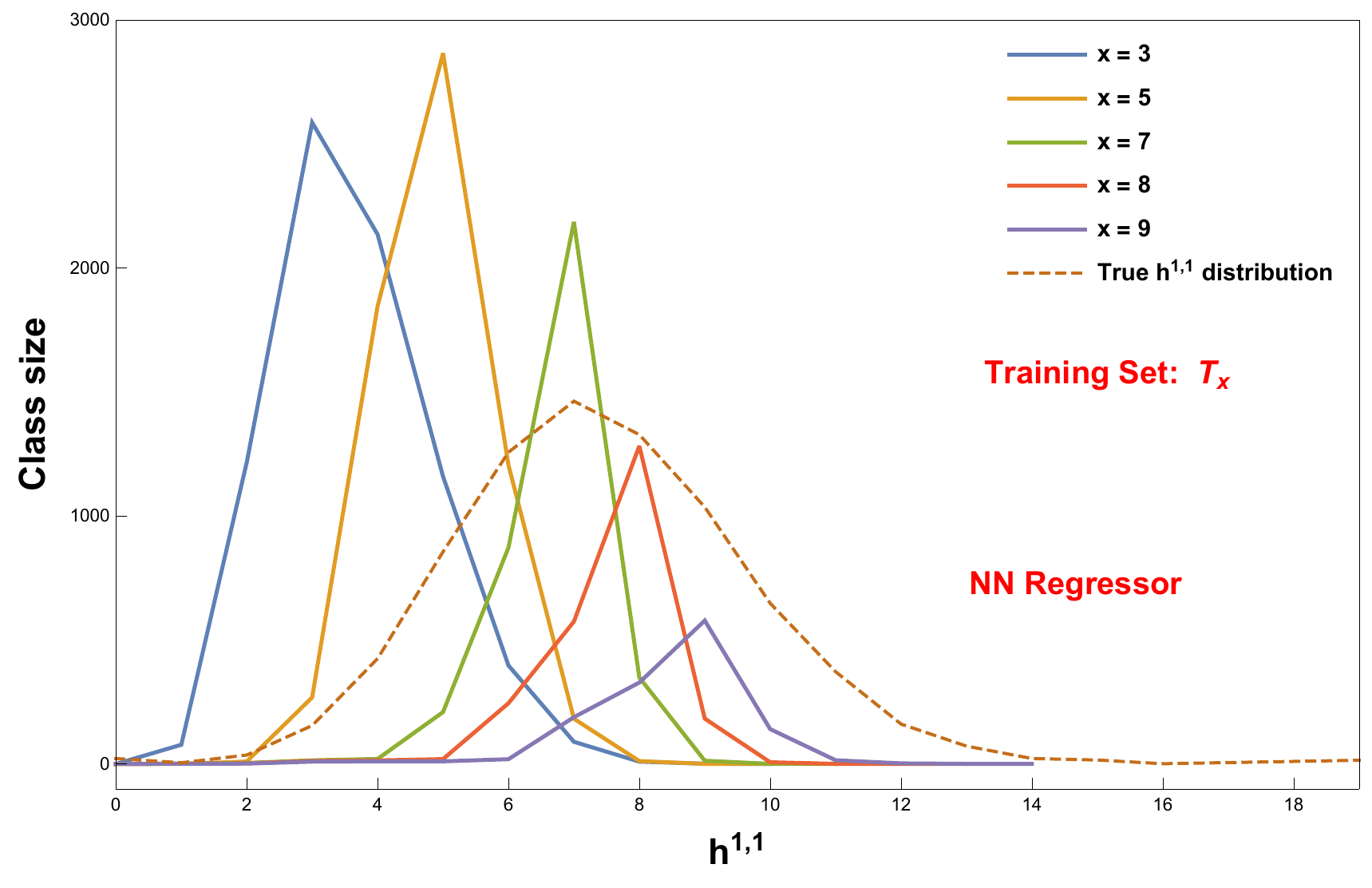}\includegraphics[width=0.49\linewidth]{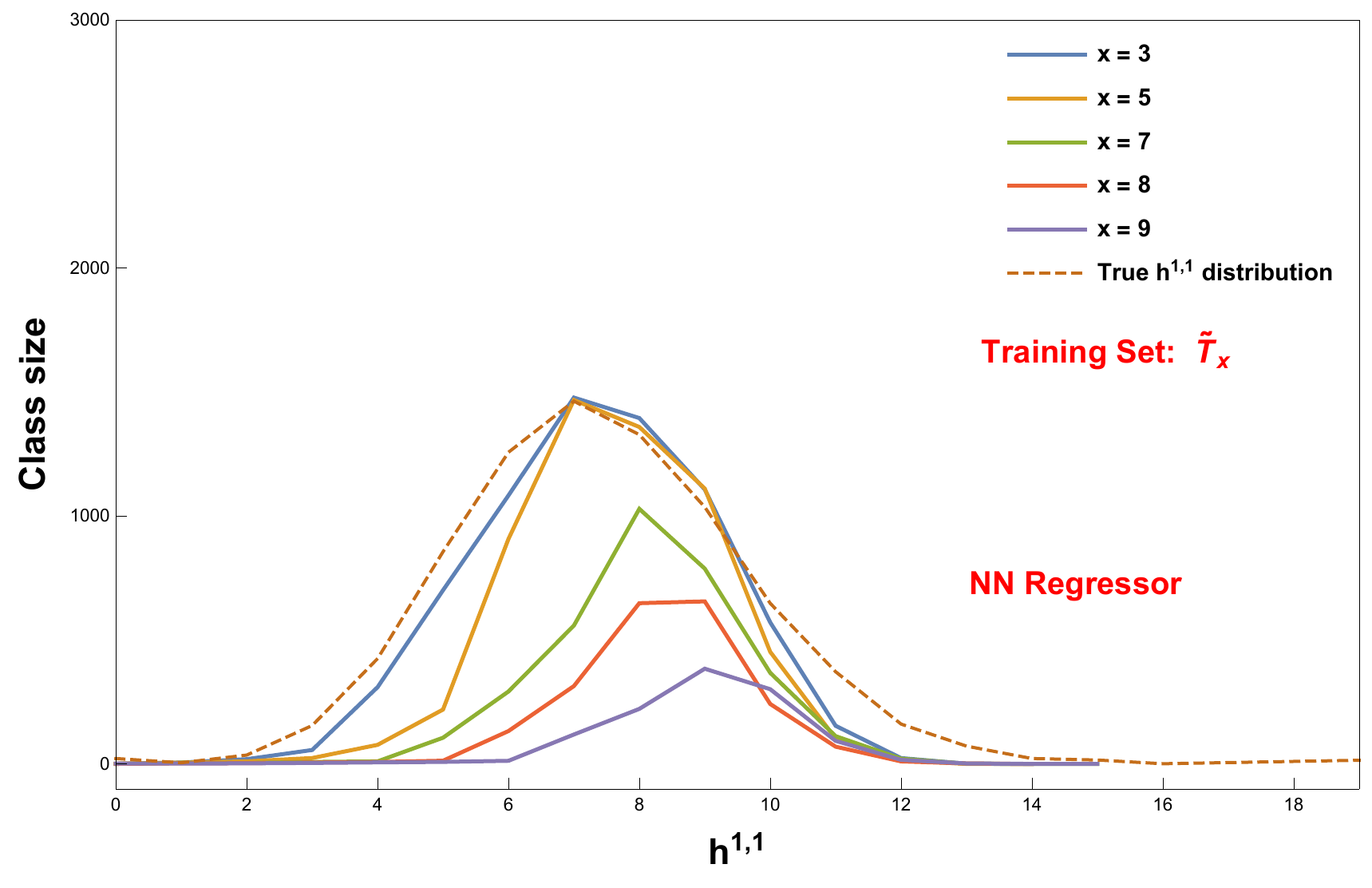}
\includegraphics[width=0.49\linewidth]{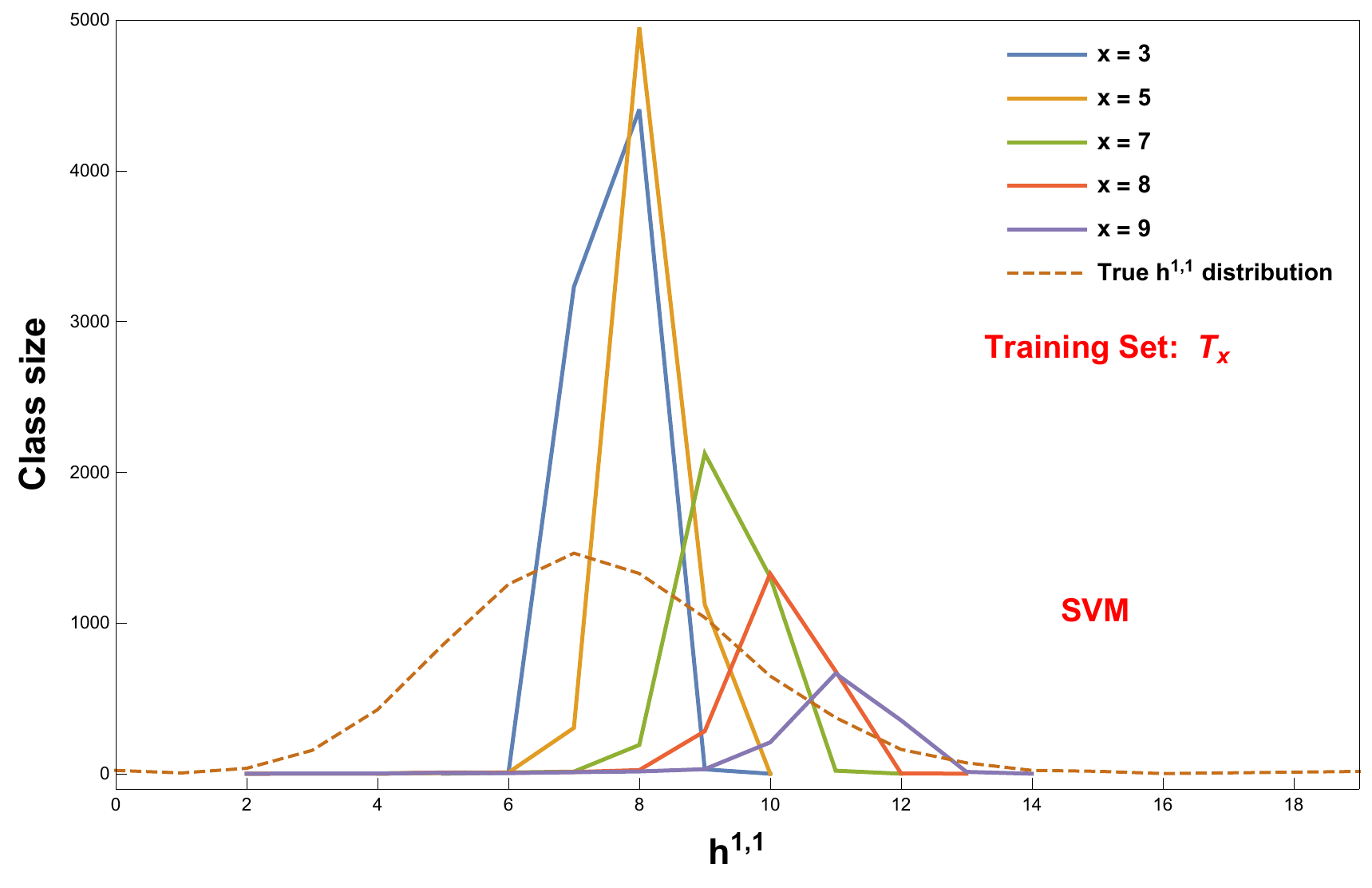}\includegraphics[width=0.49\linewidth]{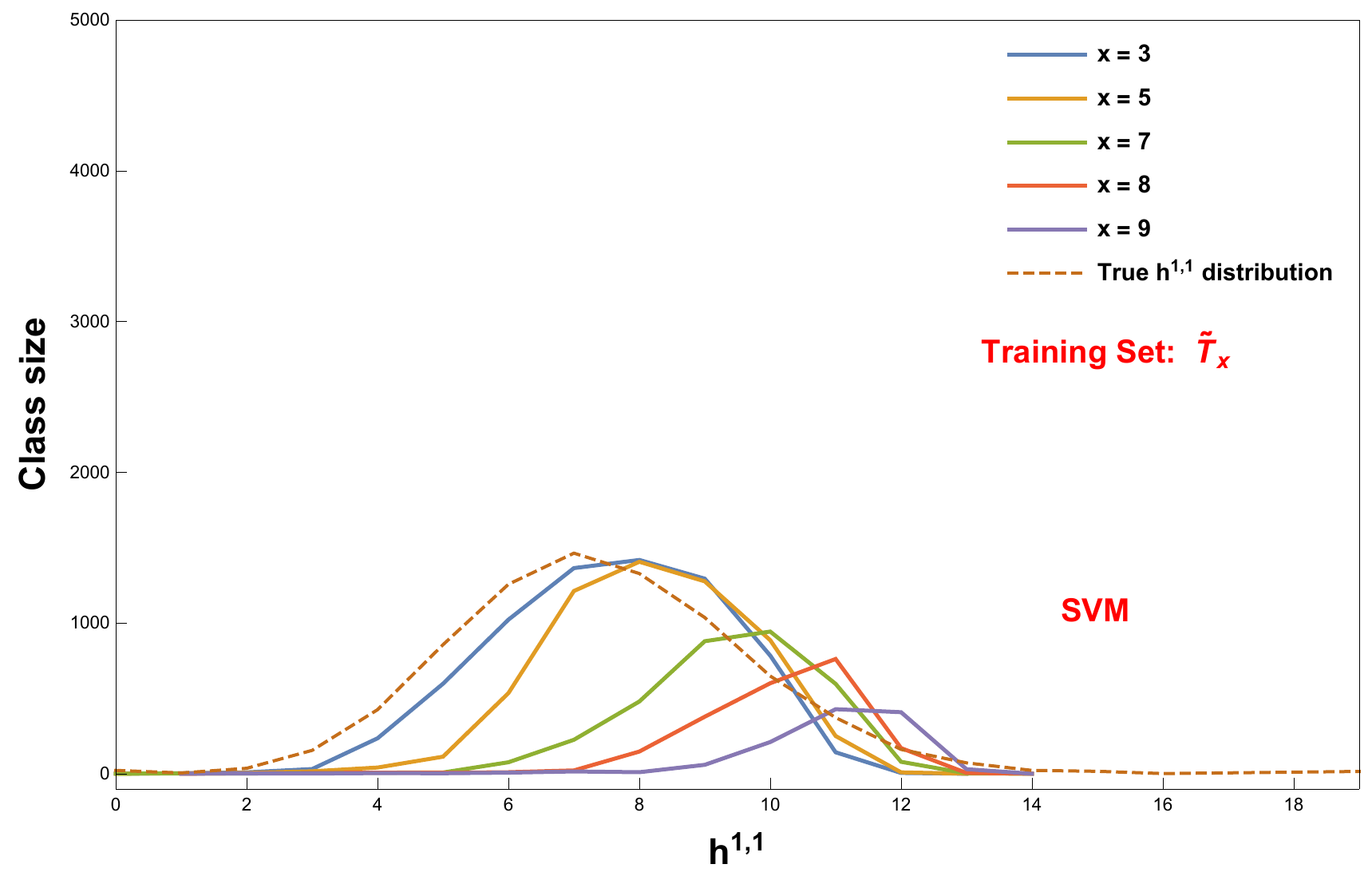}
\vspace{-5pt}
\caption{\small{Neural network and SVM predictions of $h^{1,1}$ for CICY threefolds. The first row shows predictions by the neural network classifier using the training set $\widetilde{T}_x$. The second row shows the neural network regressor predictions using training sets $T_x$ (left) and $\widetilde{T}_x$ (right). The third row shows the same for the SVM.}}
\label{fig:h11nn}
\end{figure}

\begin{figure}[!t]
\centering
\includegraphics[width=0.49\linewidth]{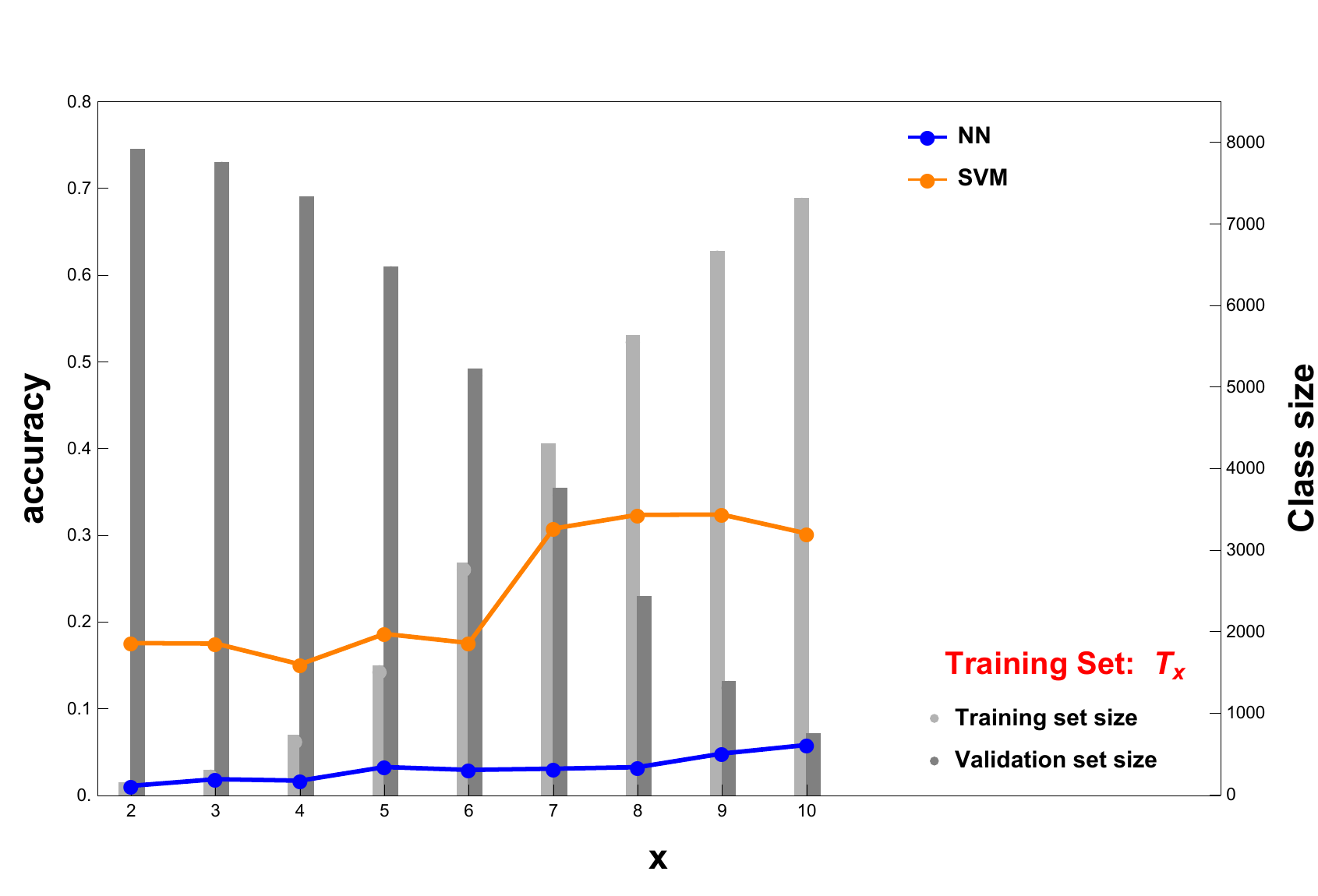} \includegraphics[width=0.49\linewidth]{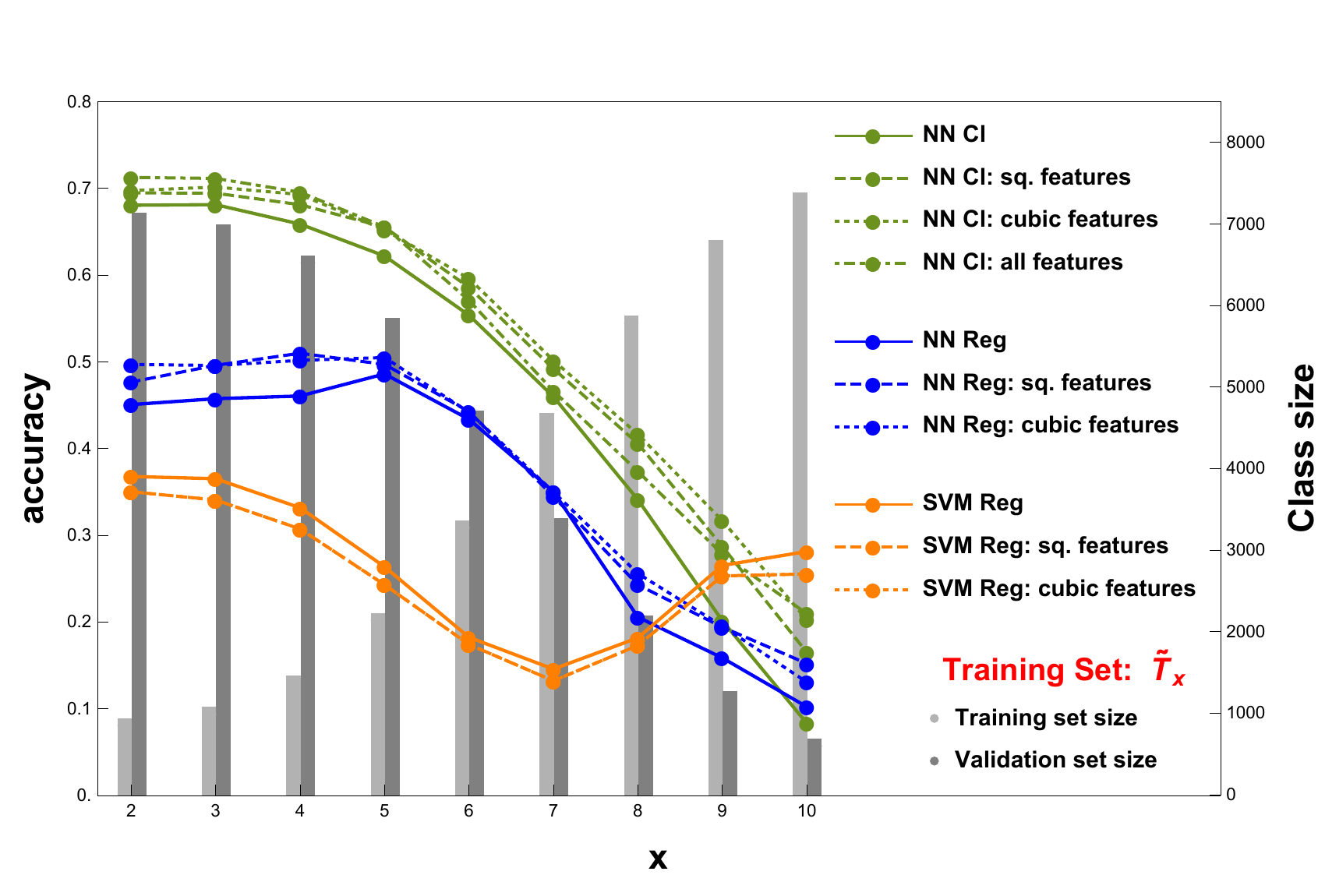}
\includegraphics[width=0.49\linewidth]{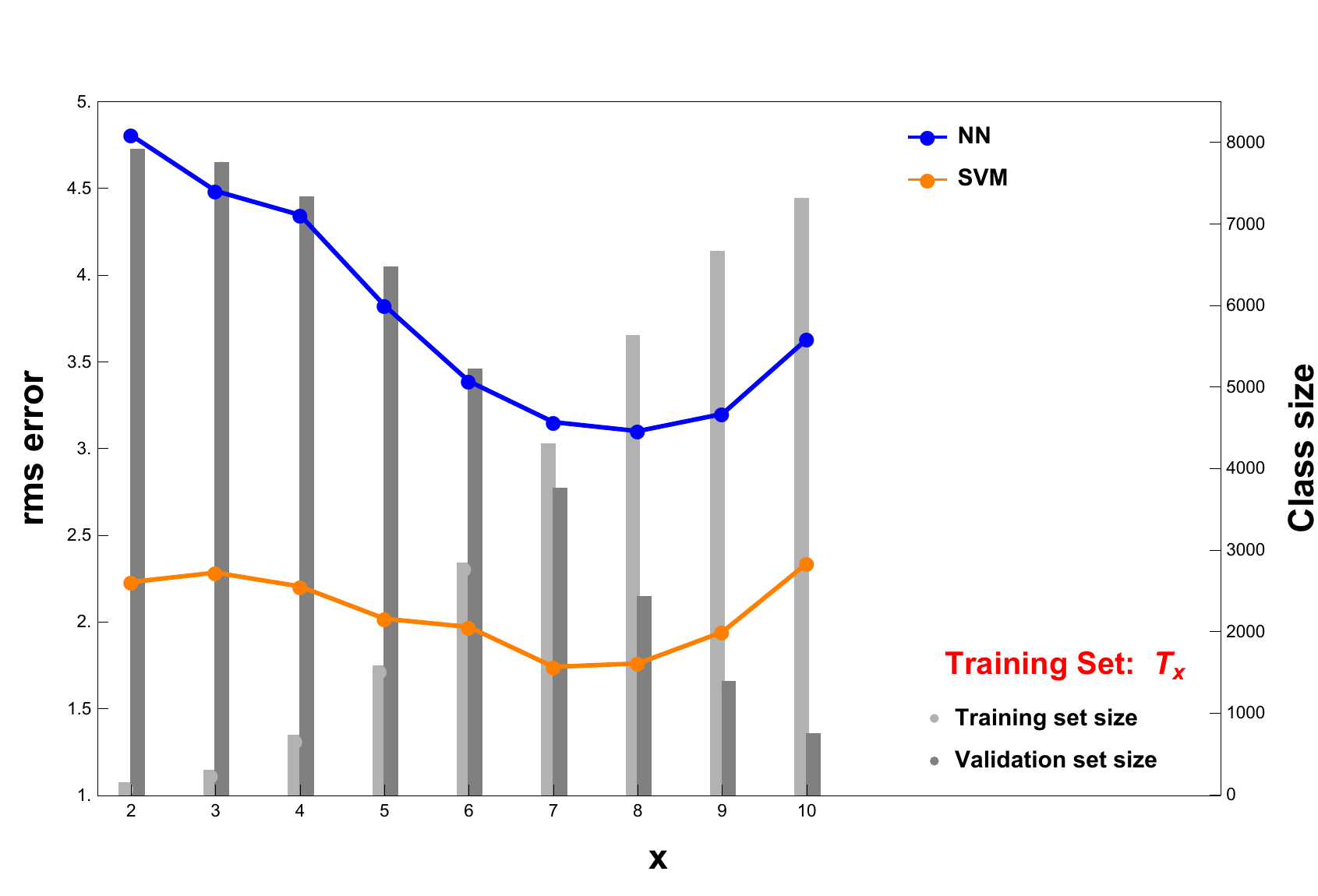} \includegraphics[width=0.49\linewidth]{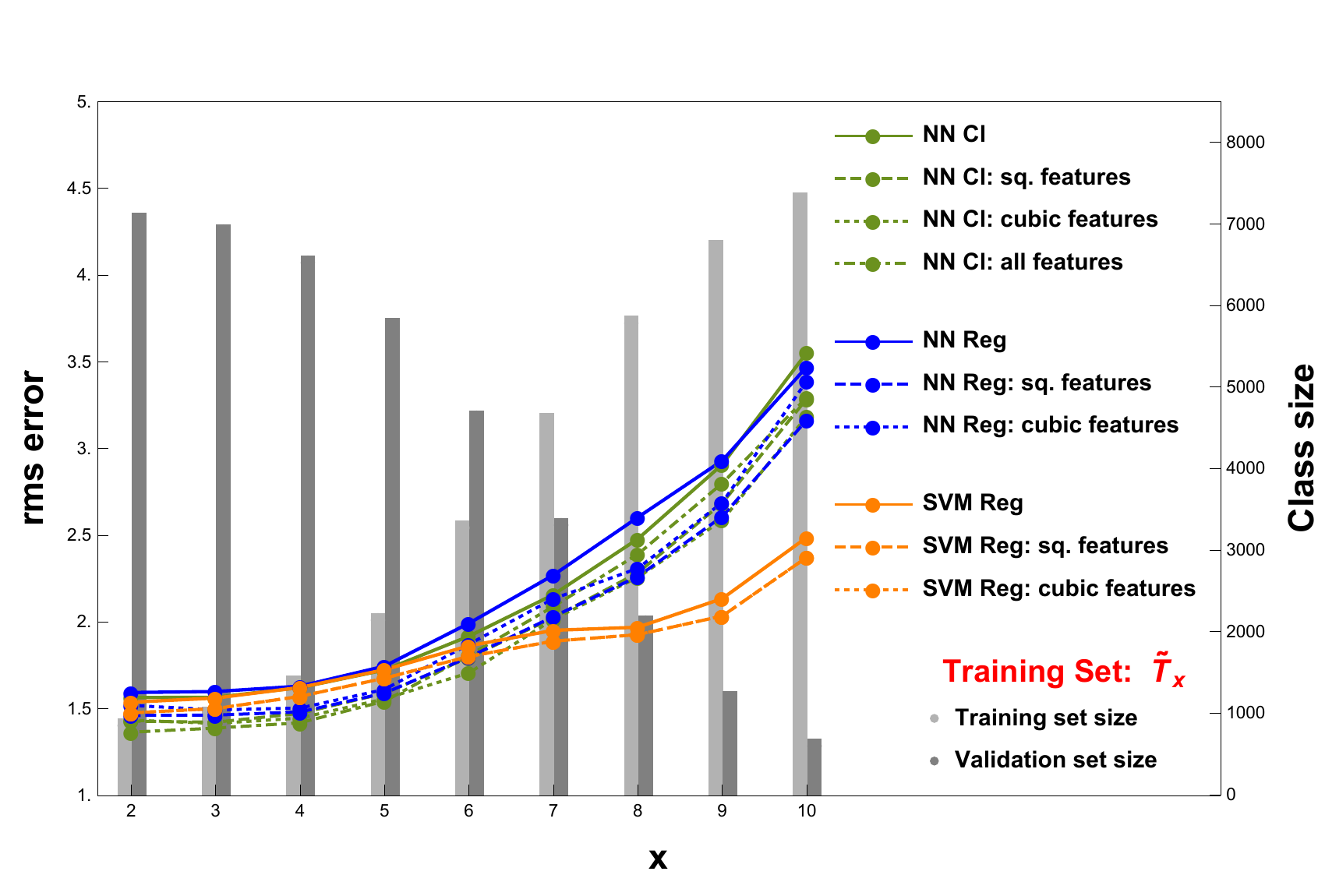}
\includegraphics[width=0.49\linewidth]{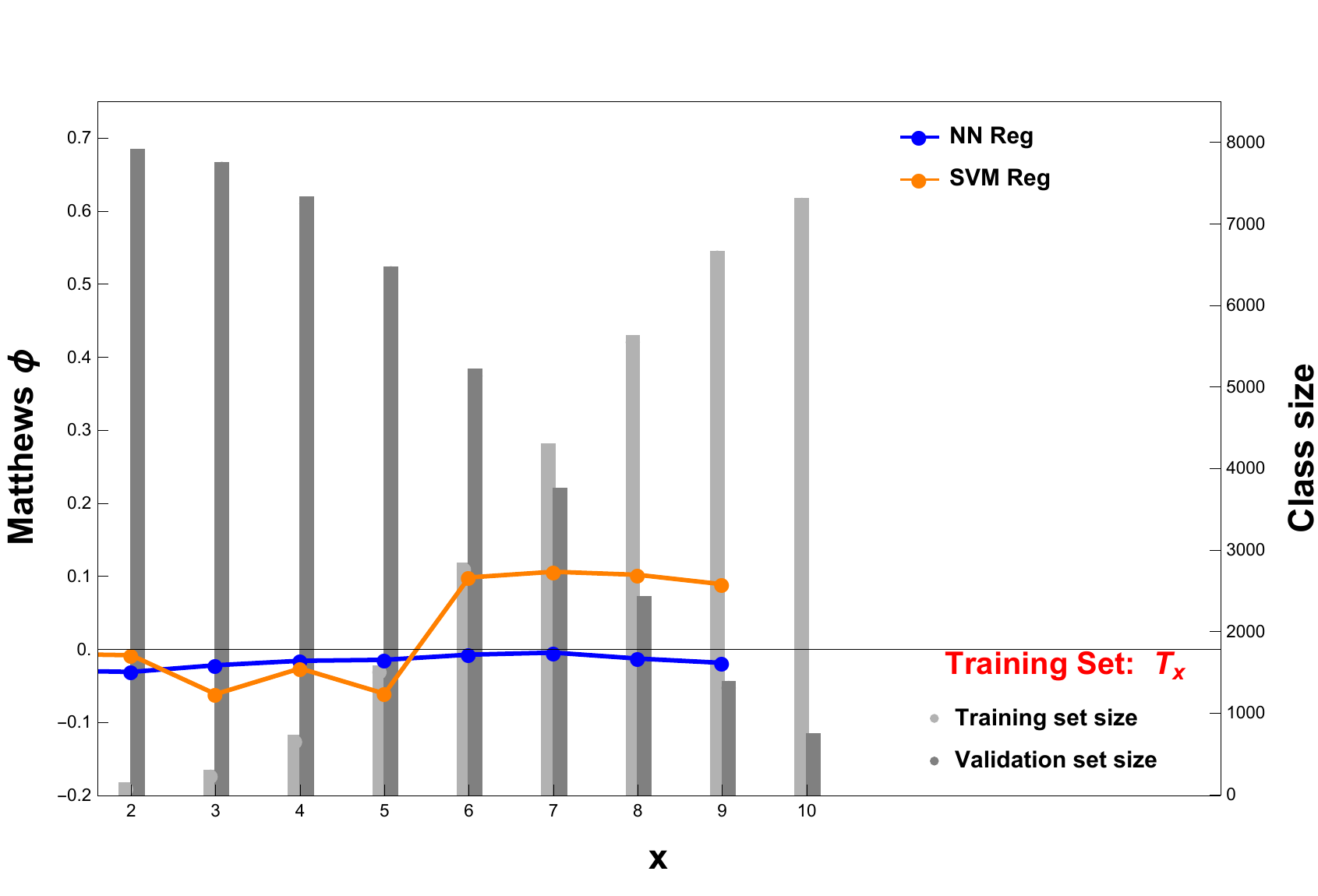} \includegraphics[width=0.49\linewidth]{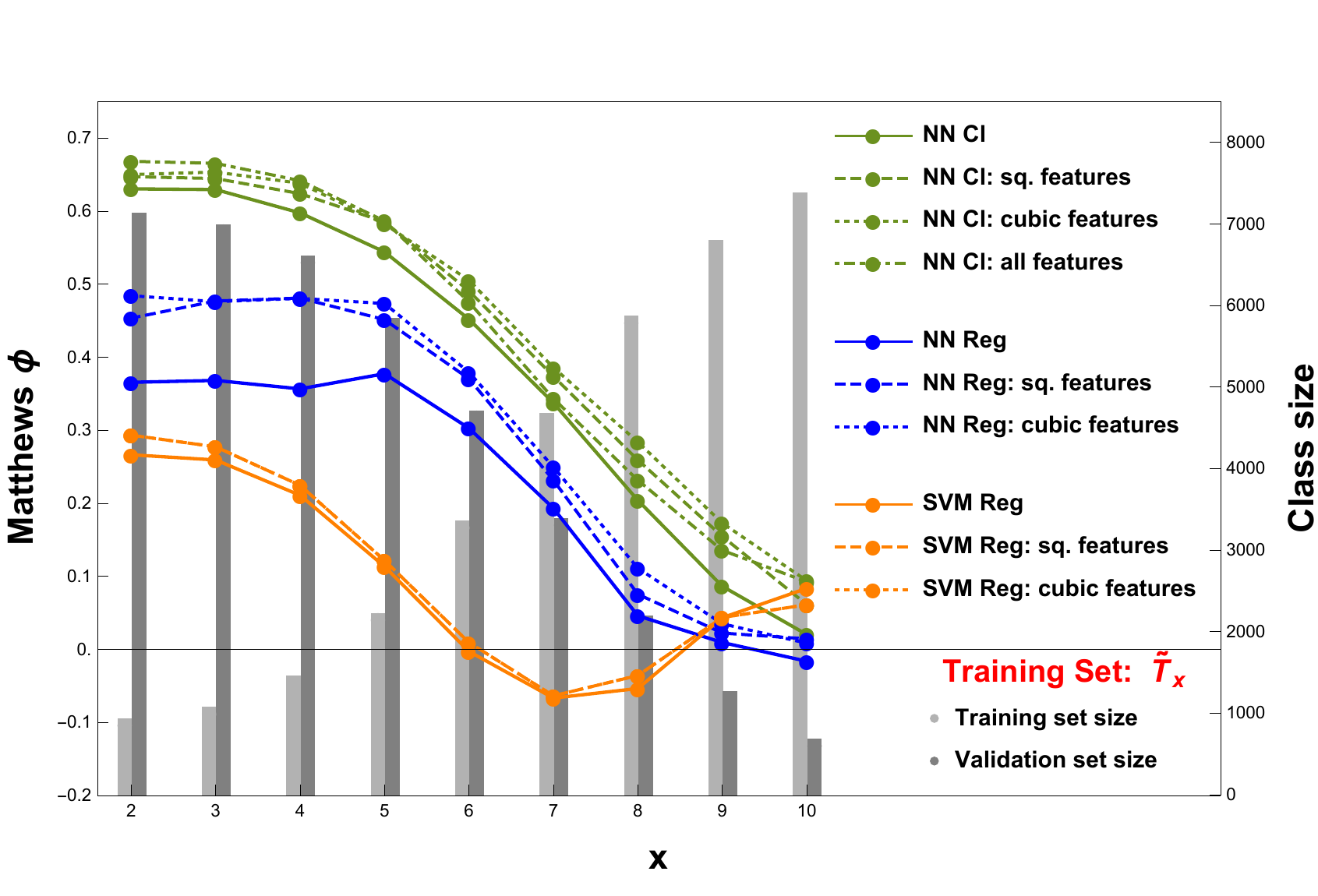}
\vspace{-5pt}
\caption{\small{Accuracy, rms error and Matthews correlation coefficient ($\phi$) of $h^{1,1}$ predictions for CICY threefolds, by the neural networks and SVM. Lighter bars represent the training set size, and darker bars, the validation set size. Figures on the left in each row correspond to experiments using the training set $T_x$, and the figures on the right correspond to experiments using the training set $\widetilde{T}_x$. For the experiments involving $\widetilde{T}_x$, we also show the effect of using the squares and cubes of the elements of the CICY configuration matrices \eqref{cicyconfmat} as input features. For the neural network classifier we also show the effect of including square and cubic features in addition to the original feature, the CICY matrices.}}
\label{fig:h11rms}
\end{figure}

The true distribution of CICY threefolds peaks at the value $h^{1,1}=7$. Figure~\ref{fig:h11nn} shows neural network and SVM predictions of this distribution. Figure~\ref{fig:h11rms} shows the accuracy, root-mean-squared (rms) errors and Matthews correlation coefficient ($\phi$) for the predictions. The left and right panels of these figures correspond to the use of the two training sets $T_x$ and $\widetilde{T}_x$ respectively, which were defined above. The neural network classifier performs rather poorly, when trained using the set $T_x$, and we exclude its predictions from Figures~\ref{fig:h11nn} and~\ref{fig:h11rms}. 

Focusing first on the experiment using the training set $T_x$, wherein we use the neural network and SVM regressors, we note that the algorithms predict a peak in the $h^{1,1}$ distribution for each value of $x$, though the position of the peak is slightly incorrect. Both the algorithms consistently overpredict the number of manifolds with low $h^{1,1}$, regardless of the parameter $x$. This is not surprising since the only data the machine has seen for training are those geometries with $h^{1,1}\le x$. This stagnates the neural network, with it eventually predicting most of the manifolds with $h^{1,1}>x$ to have  $h^{1,1}\le x$, causing the growth in the  rms error after the initial dip (Figure~\ref{fig:h11rms}). The dip itself corresponds to the better predictions as seen in the neural network plot (Figure~\ref{fig:h11nn}). From the accuracy and rms error plots (left panels in Figure~\ref{fig:h11rms}), we note that the SVM performs significantly better than the neural network, though the overall predictive powers of both the algorithms are limited. This analysis shows that the regressors are capable of predicting trends in the distribution of Hodge numbers from the limited data.

We now compare the results above with those from the experiment using the modified training set $\widetilde{T}_x$. The right panels in Figure~\ref{fig:h11nn} show the level of agreement of the predictions with the true $h^{1,1}$ distribution, demonstrating a marked improvement in the machines' predictive ability, from above. This is further evidenced by the higher accuracies and Matthews coefficient, and lower rms errors (in the right panels of Figure~\ref{fig:h11rms}). This significant enhancement of predictive ability is seemingly disproportionate to the expected gain of these algorithms (especially the neural networks) from the use of an increased number of training examples. This indicates that adding a small fraction of randomly sampled data from the list of manifolds with $h^{1,1}>x$ to the training set results in significantly improved predictions. Finally, we note that the neural networks perform better than the SVM in the domain of low $x$, and the SVM performs marginally better in the domain of high $x$. 
The accuracy, which is lower than what we report in~\cite{bhjm}, corresponds to an exactly correct identification of a manifold's $h^{1,1}$ based on an imbalanced training set.
The misidentifications follow a Gaussian profile: a prediction is more likely to be off by a little than by a lot.
Even with a simple \texttt{Mathematica} implementation, the algorithm is much better at distinguishing large from larger $h^{1,1}$.

As we have noted in Section~\ref{sec:two}, the Euler character is cubic in the elements of the configuration matrix.
It is also proportional to the difference between $h^{1,1}$ and $h^{2,1}$.
Instead of training with the elements $m_{ij}$ of the CICY configuration matrix, suppose we use $m_{ij}^2$ or $m_{ij}^3$ as inputs.\footnote{
We thank Andre Lukas for suggesting this experiment.}
We can nudge the performance slightly.
The square and cubic inputs both yield nearly the same results (Figure \ref{fig:h11rms}). 
The neural networks respond more favorably to the alternative input than the SVM.

\section{Discussion}\label{sec:disc}
The difficulty of exploring the string landscape and characterizing the vacuum space of solutions is technical.
We cannot perform detailed calculations, for instance, in Standard Model building, when the Hodge numbers are large.
Indeed, even finding all triangulations of a reflexive polytope at $h^{1,1} \ge 7$ to determine the full set of toric Calabi--Yau threefolds that are candidate geometries for superstring compactification has not been accomplished~\cite{cyd}.
A similar systematic effort for fourfolds in F-theory has not even been attempted.
As a result, we do not know how many string vacua there are and what fraction of these resemble the real world.

Supervised machine learning provides a structure to attack this class of problems in the face of incomplete data.
Studying CICY geometries, this letter suggests that the strategy to employ is to compute simple examples and a representative smattering of the harder cases.
This supplies the information that the machine requires to predict trends in the data and achieve results roughly comparable to sampling from the entire dataset.
Something similar happens when neural networks learn the hyperbolic volume of knot complements from Jones polynomials~\cite{jkp}.
The answers we obtain offer a starting point by flagging geometries that a string phenomenologist or a string theorist might find interesting.
Because the answers are not always error-free, we view this as an example of probably approximately correct learning~\cite{valiant}.

The topological invariants of CICY geometries are by now extremely well studied.
We have therefore not learned anything new about these manifolds as a result of this investigation.
The work of~\cite{he2017deep,He:2017set,He:2018jtw,bhjm} and what we report here nevertheless teaches us something profound.
The traditional methods for computing topological features of Calabi--Yau geometries --- sequence chasing, doubly exponential Gr\"obner basis algorithms, etc.\ --- may not be the most efficient way to proceed.
Machine learning responds to these queries in polynomial time.
We therefore conclude that there are better ways to calculate.

How does a machine learn?
At the most basic level, the problems we confront in computational algebraic geometry reduce to finding the (co-)kernels of integer matrices.
We have a black box that applies this process to land on useful semantics without knowing any syntax.
The central open question is to dissect the black box and translate these algorithms into something a human can understand and implement.
We aim to report progress in this endeavor in future work.
\vspace{-10pt}
\section*{Acknowledgements}
YHH thanks the Science and Technology Facilities Council, UK, for grant ST/J00037X/1, the Chinese Ministry of Education, for a Chang-Jiang Chair Professorship at NanKai University and the City of Tian-Jin for a Qian-Ren Scholarship, as well as Merton College, Oxford, for her enduring support. 
VJ is supported by the South Africa Research Chairs Initiative of the DST/NRF.
CM is supported by a Severo Ochoa Fellowship at ICMAT. We thank Dustin Cartwright, Mario Garcia-Fernandez, and Michele Cicoli for insightful comments. We also thank participants at the ICERM workshop on ``Non-linear algebra'' at Brown University and the workshop ``Machine Learning Landscape" at ICTP.  We are especially grateful to Andre Lukas for discussions. Some of the computations in this letter were carried out using the LOVELACE computing cluster at ICMAT.
\vspace{-10pt}

\bibliographystyle{JHEP} 
\bibliography{bibfile}

\end{document}